\documentclass[journal]{IEEEtran}

\usepackage{amsmath}
\usepackage{amssymb}
\usepackage{algorithm}
\usepackage{algpseudocode}
\usepackage{graphicx} 
\usepackage{multirow}
\usepackage{hhline}
\usepackage{enumitem}

\newcommand{\dd}{\text{d}}

\begin{document}
\pagestyle{plain}

\title{Volumetric Material Decomposition Using Spectral Diffusion Posterior Sampling with a Compressed Polychromatic Forward Model}
\author{Xiao~Jiang,
        Grace~J.~Gang,
        and~J.~Webster~Stayman 

\thanks{This work is supported, in part, by NIH grant R01EB030494.}
\thanks{Xiao Jiang and J.Webster Stayman are with the Department of Biomedical Engineering, Johns Hopkins University, Baltimore, MD, 20205 USA. e-mail: xjiang43@jhu.edu, web.stayman@jhu.edu}% <-this % stops a space
\thanks{Grace~J.~Gang is with the Department of Radiology, University of Pennsylvania, Philadelphia, PA, 19104 USA. e-mail: grace.j.gang@pennmedicine.upenn.edu}}

% make the title area
\maketitle

\begin{abstract}
We have previously introduced Spectral Diffusion Posterior Sampling (Spectral DPS) as a framework for accurate one-step material decomposition by integrating analytic spectral system models with priors learned from large datasets. This work extends the 2D Spectral DPS algorithm to 3D by addressing potentially limiting large-memory requirements with a pre-trained 2D diffusion model for slice-by-slice processing and a compressed polychromatic forward model to ensure accurate physical modeling. Simulation studies demonstrate that the proposed memory-efficient 3D Spectral DPS enables material decomposition of clinically significant volume sizes. Quantitative analysis reveals that Spectral DPS outperforms other deep-learning algorithms, such as InceptNet and conditional DDPM in contrast quantification, inter-slice continuity, and resolution preservation. This study establishes a foundation for advancing one-step material decomposition in volumetric spectral CT.
\end{abstract}


\section{Introduction}
\IEEEPARstart{M}{aterial} decomposition, which estimates the physical densities of basis materials, is a critical element in many spectral CT applications, such as bone density estimation, virtual monochromatic imaging, and contrast agent quantification\cite{so2021spectral,https://doi.org/10.1002/mp.14157}. However, because of the similar energy dependence of mass attenuation coefficients in typical basis materials and the often limited spectrum separation in spectral CT, material decomposition can be a highly ill-conditioned nonlinear inverse problem. Various methods have been developed to achieve fast, accurate, and low-noise material decompositions. Among those methods are one-step decomposition algorithms, which optimize data consistency often with some kind of regularization. These methods feature high decomposition accuracy and flexibility via accurate physical models and do not require geometrically matched spectral projection data. However, such approaches typically need extensive computation time/many iterations, and traditional regularization schemes can introduce artifacts such as edge blurring.

To overcome these limitations, we recently developed a physics-informed deep-learning-based one-step material decomposition algorithm, termed Spectral Diffusion Posterior Sampling (Spectral DPS)\cite{jiang2024multimaterialdecompositionusingspectral}. This method features alternating updates between a learning-based diffusion step and a model-based iterative reconstruction(MBIR) step, allowing integration of arbitrary analytic spectral system models with sophisticated priors learned from large image datasets. Both simulation and phantom studies demonstrated the ability of Spectral DPS to achieve accurate and robust material decomposition on different physical spectral system models using a one-time trained network.

Modern spectral CT scanners are capable of volumetric image reconstruction to cover a large scan area. For example, spectral cone-beam CT (CBCT) has been investigated for its potential to provide 3D quantitative information in image-guided radiotherapy and interventions \cite{Zhu_2022,https://doi.org/10.1118/1.4863598}. However, most deep learning algorithms, including the original implementation of Spectral DPS, are primarily designed for 2D imaging. Extending 2D algorithms to 3D is theoretically feasible. However, this extension faces computational challenges including large memory requirements and compute time.

In this work, we investigate a 3D Spectral DPS framework designed to operate within limited computational resources. To address memory constraints during the diffusion step, we leverage a pre-trained 2D diffusion model to process the volume slice-by-slice, rather than training a memory-intensive 3D diffusion model. Discontinuities between adjacent slices are mitigated using an additional total variation (TV) regularization along the $z$-direction. For the MBIR step, we optimize computational efficiency by calibrating a compressed polychromatic forward model, which uses significantly fewer energy bins while maintaining an accurate forward projection representation. Simulation studies demonstrate that the proposed method enables one-step volumetric material decomposition using modest computing requirements for better image quality and quantification than traditional processing and competing deep learning algorithms. 

\section{Materials and Methods}
\subsection{Compressed Polychromatic Forward Model}
\label{sec:compress}
Presume the following polychromatic forward model for an object containing $K$ basis materials:
\begin{equation}
    y=\int_0^{E_{max}} S(E)\exp\left\{-\sum_{k=1}^K q_k(E)\int_L \textbf{x}_k(\textbf{r})\dd l\right\} \dd E, 
\end{equation}

where $E$ represents the photon energy and $S(E)$ accounts for the overall spectral sensitivities including the polychromatic incident photon spectrum, additional filtration, and detector response\cite{9748122}. The terms $x_k(\textbf{x})$ and $q_k(E)$ denote the physical density and mass attenuation coefficients of the $k$-th basis material, respectively. A discretized model can be written as:
\begin{equation}
\label{eq:phyics_model}
y = \sum_{r=1}^R s_{r} \exp \left\{-\sum_{k=1}^K q_{kr} l_k\right\},\ l_k = \sum_m^M a_{m}x_{mk}
\end{equation}
where $m$ is the voxel index and $a_m$ is the projection path length through each voxel. The integral over energy has been discretized into energy bins: $\{E_r, r = 1, 2...R\}$ with equal spacing $\Delta E = E_{r+1}-E_r$. Leveraging the matrix notation from prior work\cite{9748122}, the ensemble of all projection data from a spectral CT scan can be represented compactly as:
\begin{equation}
\label{eq:phyics_model_compact}
    \overline{\textbf{y}} = \textbf{S}\exp\{-\textbf{QAx}\}
\end{equation}
Quantitative one-step material decomposition requires an accurate description of the spectrum and mass attenuation coefficients. Prior work\cite{jiang2024multimaterialdecompositionusingspectral} has used a fine spacing with more than 100 energy bins e.g. $\Delta E = 1~keV$ and $150$ bins. However, such fine discretization leads to significant memory and computational demands. To balance computational efficiency and physical accuracy, we adopt a compressed forward model (with much fewer energy bins) inspired by previous studies\cite{https://doi.org/10.1002/mp.16590}:
\begin{equation}
y_c = \sum_{r=1}^C \hat{s}_{r} \exp \left\{-\sum_{k=1}^K \hat{q}_{kr} l_k\right\},  \ C \ll R.
\end{equation}
This formulation retains the mathematical structure of \eqref{eq:phyics_model} while employing compressed energy dimensions, $C$, to enhance computational efficiency. The effective spectral sensitivity $\hat{s}_{r}$ and mass attenuation $\hat{q}_{kr}$ are found via a precalibration optimization step using the following objective:
\begin{equation}
\label{eq:objective}
\hat{s},\hat{q} = \text{argmin} \left\|\frac{y_c(s,q,l)-y}{y}\right\|^2_2 + \beta \left(\sum_{r=1}^C \hat{s}_r - \sum_{r=1}^R s_r\right)^2 .
\end{equation}
Reference projections $y$ are generated using the discretized model \eqref{eq:phyics_model} with $1~keV$. For this optimization a regular $K-$dimensional grid of density line integrals covering a clinically relevant range is used. The second term in \eqref{eq:objective} ensures photon preservation, and the regularization strength $\beta$ set to $10^6$. The objective \eqref{eq:objective} is minimized via the Adam optimizer with 50000 iterations. In each iteration, the density line integrals $l$ are jittered to prevent overfitting.

\subsection{3D Spectral Diffusion Posterior Sampling}
\subsubsection{Spectral Diffusion Posterior Sampling}
Model-based one-step material decomposition typically optimizes the following objective function:
\begin{equation}
\label{eq:mbmd}
    \mathcal{L}=\left\| \textbf{S}\exp({-\textbf{A}{\textbf{x}}}) -\textbf{y}\right\|_{\textbf{K}^{-1}}^2 - \log p(\textbf{x}) 
\end{equation}
where the first term enforces data consistency, and the second term, often referred to as prior information, incorporates regularization. However, traditional approaches usually rely on hand-crafted priors, which often fail to capture the true underlying distribution of the material densities. According to statistical theory, samples from the distribution $p(\textbf{x})$ can be generated by solving the following stochastic differential equation (SDE), starting from pure Gaussian noise:
\begin{equation}
\label{eq:reverse}
    \mathrm{d}\textbf{x} = \left\{-\frac{\beta_t}{2} \textbf{x}-\beta_t\nabla_{\textbf{x}_t} \mathrm{log} p_t({\textbf{x}_t})\right\}\mathrm{d}t + \sqrt{\beta_t}\mathrm{d}\bar{\textbf{w}} 
\end{equation}
where $\beta_t$ defines the time-dependent random noise variance and $\mathrm{d}\textbf{w}$ represents the standard Wiener process. The term $\nabla_{\textbf{x}_t} \mathrm{log} p_t({\textbf{x}_t})$, known as the score function, is typically difficult to compute analytically. Diffusion models\cite{song2020score} have been used to approximate the score by training a score network $\textbf{s}_\theta(\textbf{x}_t,t) \approx \nabla_{\textbf{x}_t} \mathrm{log} p_t({\textbf{x}_t})$. In practice, Denoising Diffusion Probabilistic Model (DDPM), discretizes the SDE into $T$ time steps. A denoising network is then trained to estimate the scaled score function via denoising score matching\cite{song2020score}: $\boldsymbol{\epsilon}_{\theta}(\textbf{x}_t,t)\approx-\sqrt{1-\bar{\alpha}_t} \ \nabla_{\textbf{x}_t}\mathrm{log} p_t({\textbf{x}_t})$ where $\bar{\alpha}_t = \prod_{i=1}^t(1-\beta_i)$. 

To solve the material decomposition problem, which minimizes the conditional probability $p({\textbf{x}|\textbf{y}})$, we replace the $p_t({\textbf{x}_t})$ with the conditional probability $p_t({\textbf{x}_t}|\textbf{y})$ and, leveraging Bayes' rule, we have:
\begin{equation}
\label{eq:dps}
\begin{split}
    \mathrm{d}\textbf{x} = &\left\{-\frac{\beta_t}{2} \textbf{x}-\beta_t\nabla_{\textbf{x}_t} \mathrm{log} p_t({\textbf{x}_t})\right\}\mathrm{d}t + \sqrt{\beta_t}\mathrm{d}\bar{\textbf{w}} \\
    &- \beta_t\nabla_{\textbf{x}_t} \mathrm{log} p_t({\textbf{y}|\textbf{x}_t})\mathrm{d}t .
\end{split} 
\end{equation} 
Combining the posterior mean approximation ($\hat{\textbf{x}}_0 = \frac{1}{\sqrt{\overline{\alpha}_t}}(\textbf{x}_t+(1-\overline{\alpha}_t)\textbf{s}_\theta(\textbf{x}_t,t))$)\cite{chung2022diffusion}, the polychromatic forward model, and several improved strategies from prior work\cite{jiang2024strategies}, yields the previously proposed Spectral Diffusion Posterior Sampling (Spectral DPS)\cite{jiang2024multimaterialdecompositionusingspectral} for one-step 2D material decomposition. This method samples from the following SDE:
\begin{equation}
\label{eq:sdps}
\begin{split}
    \mathrm{d}\textbf{x} = &\underbrace{\left\{-\frac{\beta_t}{2} \textbf{x}-\beta_t\nabla_{\textbf{x}_t} \mathrm{log} p_t({\textbf{x}_t})\right\}\mathrm{d}t + \sqrt{\beta_t}\mathrm{d}\bar{\textbf{w}}}_{Diffusion \ step} \\
    &- \underbrace{\eta\nabla_{\hat{\textbf{x}}_0} \|\textbf{S}\exp\{-\textbf{QA}\hat{\textbf{x}}_0\}-\textbf{y}\|_{\textbf{K}^{-1}}^2\mathrm{d}t}_{MBIR \ step} .
\end{split} 
\end{equation} 
Eq.~\eqref{eq:sdps} alternates between a diffusion step and an MBIR step, ensuring that the final output adheres to both the target distribution and the nonlinear spectral measurement model.

\subsubsection{Volumetric Material Decomposition Using 2D Diffusion Model and 3D Physics Model}
Previous work\cite{jiang2024multimaterialdecompositionusingspectral} demonstrated the superior performance of Spectral DPS in a 2D CT scenario. However, extension to 3D poses challenges including large memory requirements for training a 3D score network and a high computational burden for evaluating the 3D forward model due to spectral effects modeling. To address the first issue, we adopt a common strategy of training a 2D score network and processing the 3D volume slice-by-slice. To mitigate discontinuities between adjacent slices, a TV penalty is added along the longitudinal direction in the MBIR step:
\begin{equation}
\mathcal{L}=\left\| \textbf{S}\exp({-\textbf{A}{\textbf{x}}}) -\textbf{y}\right\|_{\textbf{K}^{-1}}^2 + \beta_{TV} \|\partial_z \textbf{x}\|_1 
\end{equation}
For the second issue, the compressed forward model and an ordered subset strategy are utilized to enable volumetric material decomposition with reasonable computational cost. The overall decomposition algorithm is summarized in Algorithm 1. We use 8 subsets and optimize other hyperparameters discussed in the 2D work\cite{jiang2024multimaterialdecompositionusingspectral}.

\begin{algorithm}
\caption{Volumetric Spectral DPS }
\begin{algorithmic}[1]\small
\State $\textbf{Q}_c, \textbf{S}_c$: compressed spectrum and mass attenuation
\State $T/T' = 1000/140$: training/sampling time steps
\State $\boldsymbol{\epsilon}_{\theta}$: 2D score network
\State $\hat{\textbf{x}}_0^i$: image-domain decomposition
\State
\State \# Adam optimizer 
\State $\eta=0.003, \gamma_1 = 0.9, \gamma_2 = 0.999$: step size, momentum coeff
\State
\State \# Ordered subset
\State $S=8$: number of subset
\State $\{\textbf{A}_s, \textbf{y}_s\}$: subset forward projector and spectral measurements
\State
\State \# Reverse sampling initialization
\State $\boldsymbol{\epsilon}\sim\mathcal{N}~(0,\textbf{I})$
\State $\textbf{x}_{T'} = \hat{\textbf{x}}_{T'}^i=\sqrt{\bar{\alpha}_{T'}}\hat{\textbf{x}}_0^i + \sqrt{1-\bar{\alpha}_{T'}}\boldsymbol{\epsilon}$
\State
\State \# Posterior sampling
\For{\texttt{$t = T'$ to $1$}}:
    \State \# {Diffusion step}:
    \State Loop each slice of $\textbf{x}_t$:
    \State $\ \ \ \ \textbf{z} \sim \mathcal{N}~(0,\textbf{I})$
    \State $\ \ \ \ \hat{\textbf{x}}_0 = \frac{1}{\sqrt{\bar{\alpha}_t}}(\textbf{x}_t-\sqrt{1-\bar{\alpha}_t}\boldsymbol{\epsilon}_\theta(\textbf{x}_t,t))$
    \State $\ \ \ \ \textbf{x}_{t-1}' = \frac{\sqrt{\alpha_t}(1 - \bar{\alpha}_{t-1})}{1 - \bar{\alpha}_t} \textbf{x}_t + 
    \frac{\sqrt{\bar{\alpha}_{t-1}}\beta_t} {1 - \bar{\alpha}_t} \hat{\textbf{x}}_0 + \sigma_t \textbf{z}$
    \State
    \State \# {MBIR step}:
    \State $\hat{\textbf{x}}_0' = \hat{\textbf{x}}_0$ 
    \For{\texttt{$s = 1$ to $S$}}:
        \State $\hat{\textbf{x}}_0' = \text{Adam}(\hat{\textbf{x}}_0',S\nabla_{\hat{\textbf{x}}_0'}\left\| \textbf{S}_c\exp({-\textbf{Q}_c\textbf{A}_s\hat{\textbf{x}}_0'}) -\textbf{y}_s\right\|_{\textbf{K}^{-1}}^2  + \beta_{TV} \|\partial_z \textbf{x}\|_1)$
    \EndFor
    \State $\textbf{x}_{t-1} = \textbf{x}_{t-1}' - \hat{\textbf{x}}_0 + \hat{\textbf{x}}_0'$
\EndFor
\end{algorithmic}
\end{algorithm}

\subsubsection{Diffusion Model Training}
To create a diverse dataset for training the generative model, we collected chest scans from multiple public CT datasets i.e. Lymph\cite{roth2014new}, Mayo2016\cite{https://doi.org/10.1002/mp.12345}, Kits2023\cite{10.1007/978-3-031-54806-2_1}, Covid2020\cite{Covid}. We picked 30488 slices from 370 patients for training while reserved 7086 slices from other 91 patients for validation and testing. Details on dataset processing, score network configuration, and network training can be found in \cite{jiang2024multimaterialdecompositionusingspectral}.

\subsection{Evaluation}
A two-basis 3D phantom was generated from a single patient in the Lymph dataset. The volume dimensions were $512\times512\times256$ voxels with a voxel size of $0.8\times0.8\times1.0~mm^3$. A spectral CT system was simulated using a dual-layer flat-panel detector with $1024\times384$ pixels and a pixel size of $1~mm^2$. For data generation, a circular scan with 720 projections over $360^\circ$, an exposure of 0.1~mAs/view, and 150 energy bins with $1~keV$ spacing were simulated using our projector tools\cite{jiang2025ctorch}.

We compared the proposed algorithm with image-domain decomposition (IDD), a non-diffusion material decomposition network (InceptNet)\cite{gong2020deep}, and a conditional-DDPM-based material decomposition (CDDPM)\cite{peng2024image}. InceptNet directly maps FBP-reconstructed images to material images, whereas CDDPM conditions the generation process using both FBP and material images. Both methods process 3D volumes slice-by-slice using 2D networks. All algorithms were tested on a PC (AMD Ryzen 9 5950X CPU, NVIDIA GeForce RTX 4090 GPU), leveraging a custom CUDA-accelerated toolbox for forward projection and PyTorch built-in GPU acceleration.

\section{Results}
\subsection{Hyper-parameter $C$ Selection}
\begin{figure}[h]
    \centering
    \includegraphics[width=\linewidth]{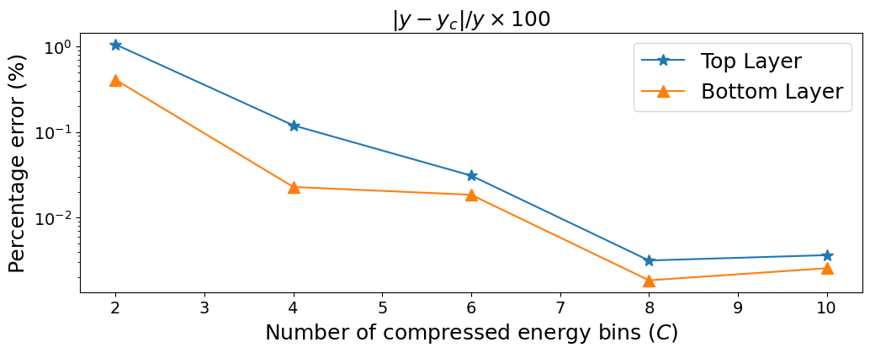}
    \caption{Percentage error of noiseless projections simulated with optimized $\hat{s}$ and $\hat{q}$ of different compressed energy bins. For both layers, $C\leq8$ can achieve an error of $<0.01\%$.}
    \label{fig:compress_dim}
\end{figure}
As described in Sec.\ref{sec:compress}, we optimize the compressed spectrum and mass attenuation ($\hat{s}, \hat{q}$) for different number of bins ($C$). The percentage error for the predicted projections on the $l$ grid is shown in Fig.\ref{fig:compress_dim}. As $C$ increases, the projection error decreases, reaching below $0.01\%$ when $C\geq8$. Further increasing $C$ to 10 yields no significant improvement. Therefore, we use $C=8$ in our compressed forward model.

\subsection{Accuracy of Compressed Forward Model}
\begin{figure}[h]
    \centering
    \includegraphics[width=0.8\linewidth]{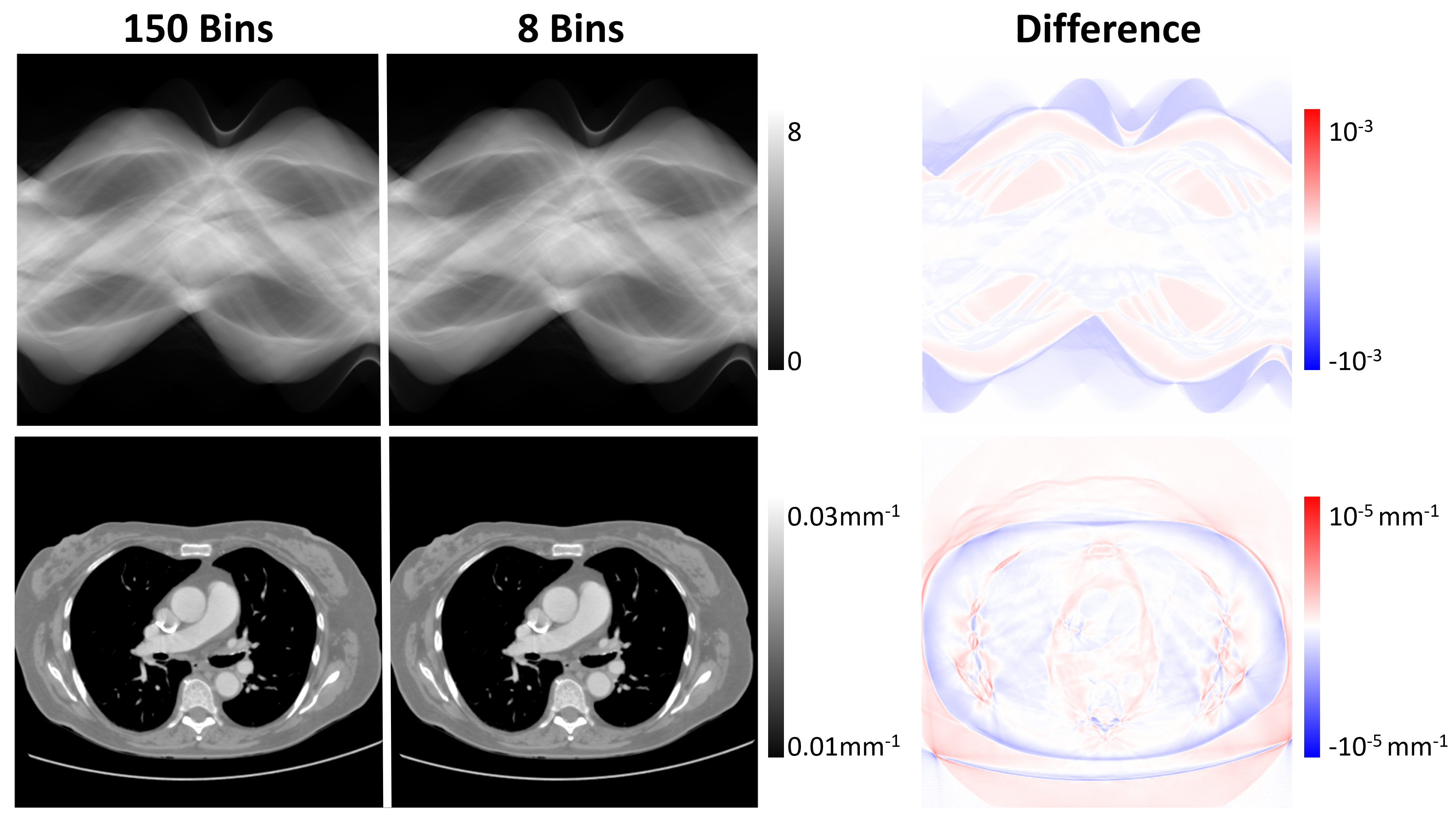}
    \caption{Sinogram and FBP images generated from noiseless projection simulated via discretized model (150 bins) and compress model (8 bins). The error maps with narrow display window indicates an ignorable error with compressed forward model.}
    \label{fig:compress_accu}
\end{figure}

\begin{figure}[h]
    \centering
    \includegraphics[width=0.8\linewidth]{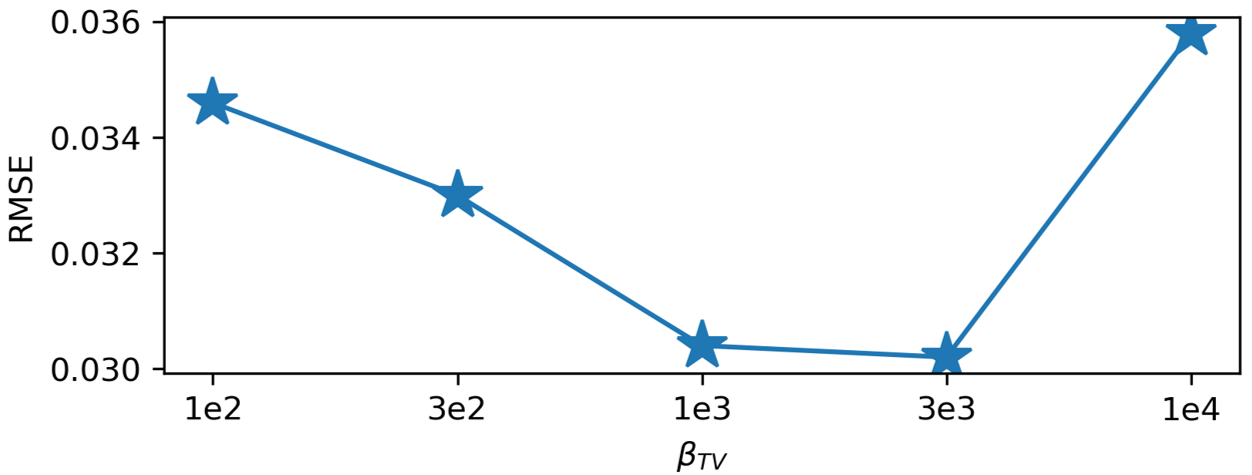}
    \caption{Reconstruction RMSE vs. regularization strength $\beta_{TV}$.}
    \label{fig:beta}
\end{figure}

\begin{figure*}[t]
    \centering
    \includegraphics[width=0.8\linewidth]{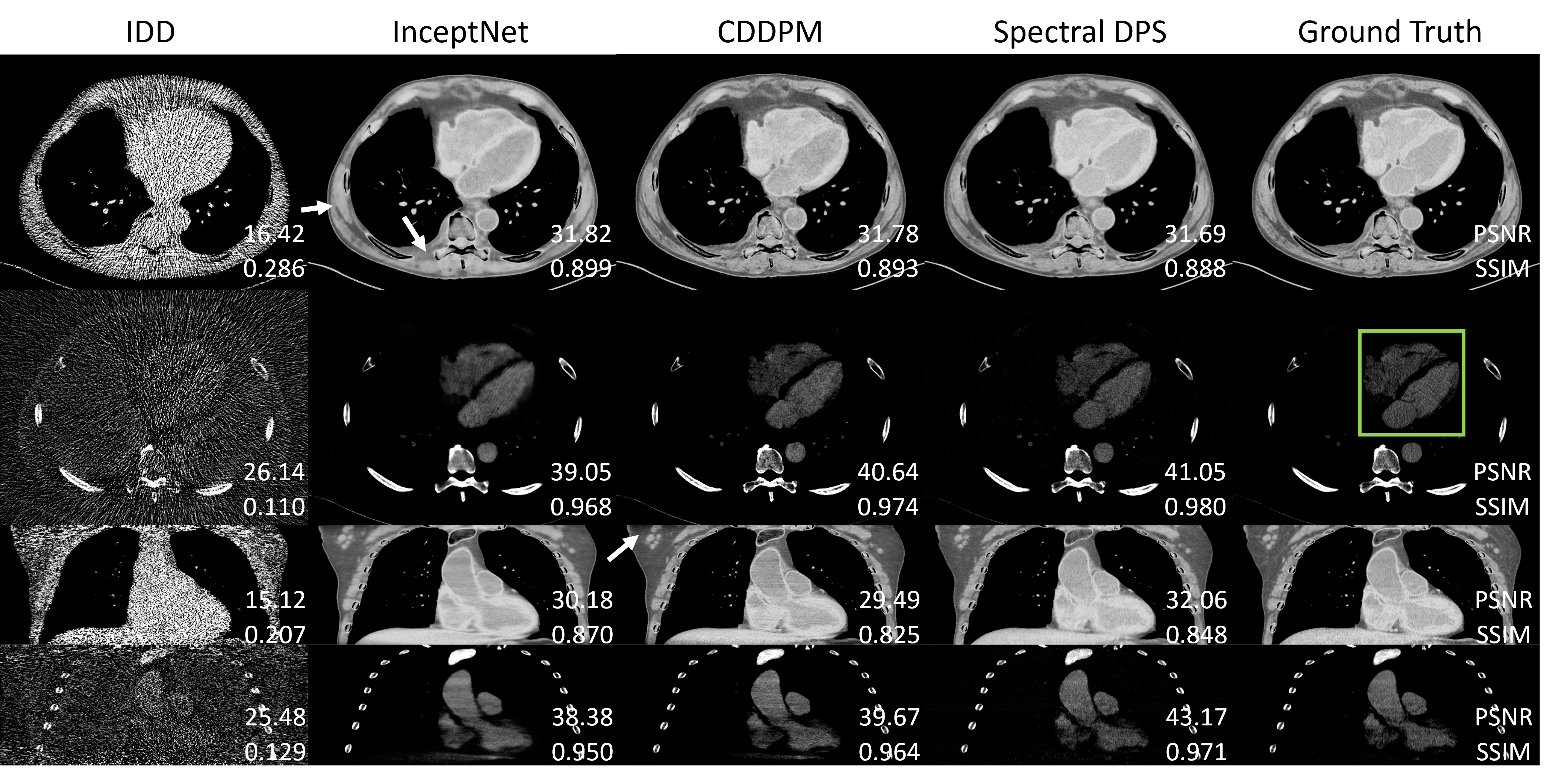}
    \caption{3D material decomposition results using different algorithms. The first and third rows are the water images, $[0.8, 1.2]g/ml$, second and last rows are the calcium images, $[0.0, 1.2]g/ml$. PSNR and SSIM are labeled on the bottom-right corner. Grean box indicates the cardiac ROI for contrast quantification analysis.}
    \label{fig:full}
\end{figure*}
We further validate the compressed forward model on a digital phantom. Fig.~\ref{fig:compress_accu} shows sinogram data and corresponding FBP images for noiseless projections computed using the 150 bin discretized model and the 8 bin compressed model. The error map, displayed with a narrow window setting, shows only small errors for the compressed model. A small background bias is observed in the reconstructed image, however the 8-bin model achieves an error of less than $0.1\%$ within the patient.

\subsection{Volumetric Material Decomposition}
Fig.~\ref{fig:beta} shows the RMSE of the entire reconstructed volume as a function TV regularization strength ($\beta_{TV}$). The optimal $\beta_{TV}$ was determined to be $3 \times 10^3$. Fig.~\ref{fig:full} compares the results of different algorithms. IDD exhibits significant noise and bias due to the ill-conditioned and linearized model, while InceptNet effectively reduces noise and improves accuracy. However, despite achieving comparable PSNR and SSIM to diffusion-based algorithms (CDDPM and Spectral DPS), InceptNet shows noticeable edge blurring, limiting soft tissue detail visualization. In contrast, diffusion-based algorithms can learn the underlying image distribution, and exhibited better resolution preservation. In the axial images, CDDPM and Spectral DPS show comparable performance on visual inspection and in quantitative metrics. However, Spectral DPS significantly outperforms CDDPM in mitigating cone-beam artifacts and adjacent slice discontinuities. 

\begin{figure}
    \centering
    \includegraphics[width=\linewidth]{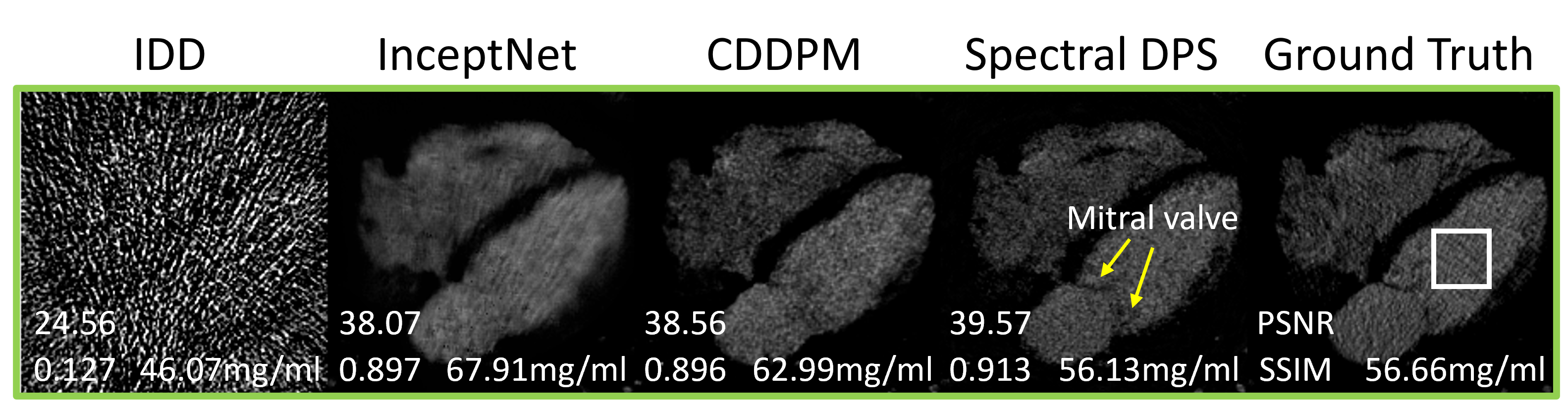}
    \caption{Cardiac ROI in calcium images. The bright signal comes from the contrast enhancement, and the mean density (bottom right) within the white box ($20\times20$ pixels$^2$) are computed to evaluate the contrast quantification accuracy.}
    \label{fig:cardiac}
\end{figure}

Contrast quantification, though not explicitly included in this study, can be inferred from the cardiovascular region visible as a low-density area in the calcium images. Fig.~\ref{fig:cardiac} displays zoomed-in images of the cardiac ROI. Spectral DPS achieved the best overall SSIM and PSNR. In the uniform box area, the mean density errors for InceptNet, CDDPM, and Spectral DPS are $19.86\%, 11.17\%,$ and $0.94\%$ error, respectively. Spectral DPS provides the clearest depiction of the mitral valve, demonstrating its superior decomposition accuracy in contrast-enhanced regions.

\subsection{Computational Efficiency}
Table \ref{tab:compute} compares the computational cost of different deep learning algorithms. InceptNet, with its end-to-end inference approach, demonstrates the lowest computational cost, requiring only 2.8 seconds and 5.6 GB of memory to process the 3D volume. Diffusion-based algorithms, which rely on iterative sampling, typically have the highest computational burden, e.g., CDDPM has a runtime of 7722 seconds and a memory requirement of 6.8 GB. The proposed algorithm, while still involving computationally intensive projection, achieves a significant reduction in both runtime (2680 seconds) and memory usage (15.9 GB) compared to the 150-bin version (4020 seconds and ~108.2 GB) by leveraging the compressed forward model. This demonstrates the improved efficiency and practical applicability.
\begin{table}[h]
    \centering
    \footnotesize
    \begin{tabular}{|c|c|c|c|c|}
        \hline
        Method     & Incept & CDDPM & Proposed(150 bin) & Proposed\\ 
        \hline
        Time(s)    &  2.8s  & 7722  & $4020^*$ & 2680\\
        \hline
        Memory(GB) &  5.6   & 6.8   & $108.2^*$  & 15.9\\
        \hline
    \end{tabular}
    \normalsize
    \caption{Computational cost of deep learning algorithms. \\${}^*$The 150 bin values are estimated by linear extrapolation of a thin volume decomposition since the memory requirements are too large for practical implementation on our system.}
    \label{tab:compute}
\end{table}
\section{Conclusion and Discussion}
In this work, we extend the 2D Spectral DPS to a 3D scenario. By leveraging a pre-trained 2D diffusion model and a compressed polychromatic forward model, Spectral DPS enables volumetric material decomposition on a single computer and GPU. Comparative studies demonstrate that Spectral DPS outperforms other deep learning algorithms in terms of accurate contrast quantification and inter-slice continuity, showcasing its capability for precise 3D material decomposition.

Future work will focus on strategies to further accelerate the computation, such as employing faster reverse SDE solvers and high-quality diffusion initialization. Moreover, application to physical data is a topics of ongoing studies. We expect that scatter correction is likely to be another major challenge for accurate volumetric material decomposition. Both classic scatter correction methods and advanced approaches\cite{lorenzon2024joint} that integrate scatter into the physical forward model for joint estimation will be investigated. We expect that this work and those improvements will lay the foundations for application of 3D spectral DPS in clinical systems.

\bibliography{report} % bibliography data in report.bib
\bibliographystyle{IEEEref}

\end{document}